\documentclass{ptptex}
\usepackage{graphicx}
\usepackage{color}
\usepackage{hyperref}

\newcommand{\bfR}{{\bf R}}
\newcommand{\bfRp}{{\bf R}'}
\newcommand{\bfr}{{\bf r}}
\newcommand{\bfrp}{{\bf r}'}

\begin{document}
\title{Quantum Monte Carlo Approaches to Nuclear and Atomic Physics}
\date{today}
\author{J. Carlson,$^1$ Stefano Gandolfi$^1$ and Alexandros Gezerlis$^{2,3}$}
\inst{
$^1$Theoretical Division, Los Alamos National Laboratory\\
Los Alamos, NM 87545, USA \\
$^2$ExtreMe Matter Institute EMMI,
GSI Helmholtzzentrum f\"ur Schwerionenforschung GmbH, 64291 Darmstadt, Germany\\
$^3$Institut f\"ur Kernphysik,
Technische Universit\"at Darmstadt, 64289 Darmstadt, Germany}

\abst{
Quantum Monte Carlo methods have proven to be  valuable in the
study of strongly correlated quantum systems, particularly nuclear physics
and cold atomic gases. Historically, such ab initio simulations have been used to study
properties of light nuclei, including spectra and form factors, low-energy
scattering, and high-momentum properties including inclusive scattering and
one- and two-body momentum distributions.  More recently they have been
used to study the properties of homogeneous and inhomogeneous neutron matter
and cold atomic gases.  There are close analogies between these seemingly
diverse systems, including the equation of state, superfluid pairing, and 
linear response to external probes.  In this paper, we compare and contrast results
found in nuclear and cold atom physics. We show updated lattice results for the 
energy of the homogeneous unitary Fermi gas and comparisons with neutron matter,
as well as for the dependence of the cold atom energy
on the mass ratio between paired particles, which yields insights on the structure of the ground state. We also provide new lattice and continuum 
results for the harmonically trapped unitary gas, again comparing neutron matter and cold atoms.
}

\maketitle

\section{Introduction}

   Quantum Monte Carlo methods\cite{kalos1962,kalos1974} have been proven very valuable in studying a host of strongly correlated quantum systems, including solid and 
liquid Helium,\cite{ceperley1977,whitlock1979,kalos1981,ceperley1995} electronic
systems\cite{ceperley1980,ceperley1989}, light nuclei\cite{carlson1987green,carlson1988alpha,pieper2001realistic} and
more recently neutron matter\cite{carlson2003quantum, gezerlis2008strongly, gandolfi2009quantum,gezerlis2010low,gandolfi2011maximum} and cold atomic Fermi gases.\cite{carlson2003superfluid,chang2004quantum,astrakharchik2004,giorgini2008} Many of these studies are formulated in the continuum to be able to describe the short-range repulsion between,
for example, helium atoms, nucleons, or electrons. For low-density neutron matter and cold Fermi gases,
where the dominant interaction is a relatively short-range attraction, lattice
methods have also proven valuable.\cite{bulgac2006,lee2006,carlson2011auxiliary,abe2009from}  In this paper we describe applications of Quantum Monte Carlo methods to the equation of state, superfluid pairing gap, and related properties of neutron matter and cold atomic gases.

   We concentrate on the zero temperature properties of cold atoms and neutron matter.  Cold Fermi atoms have a deceptively simple, essentially 
zero-range interaction, $H = \sum_i -\hbar/2m \nabla^2_i + \sum_{i,j} V_0 \delta ({\bf r}_i - {\bf r}_j)$, the strength of which ($V_0$) can be tuned to produce a very rich set of physics 
described by a relatively small set of universal parameters.  As the strength of
the (attractive) interaction $V_0$ increases, one goes from a weak BCS pairing
regime to the BEC regime of strongly bound pairs. Many simulations and experiments\cite{giorgini2008} are performed near and at the unitary limit, which is where the two-body system produces
a nearly zero-energy bound state.

    The $s$-wave interaction between two neutrons is also very attractive, nearly producing a bound state.  For very dilute neutron matter, the neutron matter and cold atom equations-of-state should be very similar as a function of the product of the fermi momentum $k_F \equiv ( 3 \pi^2 \rho)^{1/3}$ times the magnitude $a$ of the scattering length.  The $s$-wave interaction between neutrons also has a significant
effective range which causes the equation of state of neutron matter and cold atoms to diverge at relatively modest densities. The effective range of the interaction also affects the pairing gap, etc. It may be possible to use
narrow Feshbach resonances to more directly mimic neutron matter and study the
dependence of the equation of state on the effective range.\cite{marcelis2008total}  There are, of course, additional $p-$wave interactions between neutrons, though these are relatively modest at low densities.

\section{Monte Carlo Methods}

   Monte Carlo methods have proven quite effective in dealing with 
strongly correlated quantum systems.
They are most efficient when they incorporate as much knowledge 
of the physical system to be studied as possible.  
The zero-temperature Diffusion Monte Carlo (DMC) and 
Auxiliary Field Monte Carlo (AFMC) methods we employ use Monte Carlo
to propagate a trial wave function to the true ground state of quantum systems:
\begin{equation}
| \Psi_0 \rangle \ = \ \exp [-H \tau ] \ | \Psi_T \rangle \ = \ \prod_{i=1}^N \ \exp [ -H (\tau/N)  ] \ | \Psi_T \rangle ,
\end{equation}
where the imaginary time propagation is split into small imaginary time steps $\delta \tau = \tau / N$.
For small $\delta \tau$ the propagator can be evaluated accurately in terms of the two-body propagator:
\begin{eqnarray}
\langle \bfRp \ | \ \exp [ - H \delta\tau ]\  | \ \bfR \rangle & = & \langle \bfRp \ | \prod_i \exp [ - h_{i}^0 \delta \tau ] \ \ {\cal S} \prod_{i<j} \frac{
g_{ij} (\bfrp_{ij}, \bfr_{ij} ) }{g_{ij}^0 (\bfrp_{ij}, \bfr_{ij})} \  | \bfR \rangle \nonumber \\
g_{ij} (\bfrp_{ij}, \bfr_{ij} ) \ & = & \langle  \bfrp |  \ \exp [ - h_{ij} \delta \tau ] \ | \bfr \rangle,
\end{eqnarray}
where $g_{ij}$ and $g_{ij}^0$ are the interacting and free two-particle propagators respectively,
determined by the eigenstates of the interacting and free two-particle Hamiltonian: $h_{ij} = h_{ij}^0 + v(r_{ij})$ 
and $h_{ij}^0 = - \frac{\hbar^2}{m} \nabla_{ij}^2$. The factor $\exp [ - h_i^0 \delta \tau ]$ is the
free single particle propagator, a simple gaussian, and  $\bfR$ is a 3$N$-dimensional vector containing
the coordinate-space positions of all the particles.

Diffusion Monte Carlo and Auxiliary Field Monte Carlo operate in different spaces:
DMC performs the simulations in coordinate space, Monte Carlo methods are used to sample the
spatial integrals.  AFMC calculations are carried out on a lattice, and can be viewed as the evolution
of single-particle orbitals in imaginary time.  In these lattice calculations the single-particle orbitals
can be transformed between coordinate and momentum space through fast fourier transforms. In AFMC 
the Monte Carlo  is incorporated by evaluating the potential matrix
elements in terms of fluctuating auxiliary fields.\cite{carlson2011auxiliary}  

In either case, it is possible to use a very accurate short-time two-body propagator.
In the continuum DMC simulations and in the limit of zero-range interactions the propagator at unitarity takes the simple form\cite{carlson2003dilute}:
\begin{equation}
g_{ij} (\bfrp_{ij}, \bfr_{ij} )  \ = \ g_{ij}^0 (\bfrp_{ij}, \bfr_{ij} ) +  \frac{ \sqrt{ m \pi / (\hbar^2 \delta \tau)}}{4 \pi^2 r_{ij} r'_{ij}} \exp [ - m / (\hbar^2 \delta \tau) ({r_{ij}}^2 + {r'_{ij}}^2)/4],
\end{equation}
where the correction to the free particle propagator arises from the pair propagating to the same point
and then diffusion plus multiple scattering terms.
In actual simulations typically an analytic potential is used and
a short-time approximation used for the propagator:
\begin{equation}
g_{ij} (\bfrp_{ij}, \bfr_{ij} )  \ = \exp [ - V(r'_{ij}) \delta \tau / 2 ] \ \ g_{ij}^0(\bfrp_{ij}, \bfr_{ij} )  \exp [ - V(r_{ij}) \delta \tau / 2]. 
\label{eq:shorttimedmc}
\end{equation}
This expression is accurate to order $(\delta \tau)^2$. 

For lattice calculations, the simplest propagator used is for an on-site attractive interaction and either a
Hubbard-like hopping Hamiltonian or a $k^2/(2m)$ kinetic term in an expression analogous to Eq. \ref{eq:shorttimedmc}:
\begin{equation}
g_{ij} (\bfrp_{ij}, \bfr_{ij} )  \ = \exp [ - T \delta \tau / 2] \ \exp [ - V \delta \tau ] \   \exp [ - T \delta \tau/2 ]. 
\end{equation}
The potential is evaluated by Monte Carlo sampling of auxiliary fields and the kinetic energy exactly through
the use of fast fourier transforms.\cite{bulgac2006}  This simplified interaction yields a finite
range of the order of the lattice spacing.  One can remove this residual effective range by 
altering either the kinetic term $T$, adding higher-order momentum terms in the kinetic energy,
or by introducing additional auxiliary fields to give the correct low-energy two-body spectrum.\cite{endres2011, endres2012}  Lattice methods cannot be fully galilean invariant as there is a lattice cutoff at 
the high-momentum scale, and simple implementations of the improved actions can have similar effects
to a finite effective range for pairs with non-zero momentum.\cite{werner2012}.  
These corrections typically vanish in the limit of large lattices.

The AFMC lattice simulations have the great advantage that they do not suffer from a sign
problem for purely attractive two-body interactions.  The up and down spin evolution can be factorized
in AFMC and for unpolarized systems the spin up and spin down determinants are real and equal and the product
is positive. Therefore, the Monte Carlo results should be exact within
statistical errors if we can perform simulations that are sufficiently dilute, for sufficiently large number of
particles, and at low-enough  temperatures (large $\tau$). 

The DMC continuum simulations, in contrast, suffer from a sign problem as the spin up and spin 
down determinants are independent and the product can be positive or negative. The overlap of the
propagated wave function with the trial function has a statistical error that grows with respect to the
propagation time and the number of particles. The fixed-node algorithm, though, provides an accurate
upper bound to the energy by requiring the paths in the evolution not to cross planes where the 
trial wave function is zero. This can be proved to provide an upper bound to the true ground state
energy, and hence the trial wave function can be optimized as the one that produces the lowest energy.
The advantage of the DMC simulations is that the fixed-node algorithm is equally applicable to polarized
or unpolarized systems, to systems with unequal masses, etc., where the AFMC method also
suffers from a sign problem.  Similar constrained-path algorithms
exist for AFMC methods,\cite{zhang1995constrained,zhang1997constrained}
but do not typically provide upper bounds.

In both DMC and AFMC simulations it is very important to use good trial functions in strongly correlated
ground state calculations.  In DMC the fixed-node results are quite accurate if a good trial function
of BCS type is employed, as we shall discuss later.  In AFMC the results should be correct for any
trial function that has a finite overlap with the true ground state.  However the statistical errors are dramatically
reduced if one uses a BCS trial function for the trial state, as described in Ref. \citenum{carlson2011auxiliary}.
It is also extremely valuable to use a branching random walk algorithm to limit statistical errors in
both the DMC simulations, as traditionally done, and in the AFMC simulations.\cite{zhang1995constrained,zhang1997constrained}  In zero-temperature calculations
we are trying to reach the eigenstate of the transfer matrix $\exp [ -H \tau ]$, the branching random walk
algorithm is a Markov chain algorithm, that is, only the most recent history of the path is required for
performing the next step. Consequently one can iterate to very low temperatures/large imaginary times compared to other
algorithms.

For quantities other than the energy, we often  evaluate matrix elements of the form:
\begin{equation}
\langle O (\tau) \rangle = \frac{\langle \Psi_T | O \exp [ - H \tau ] | \Psi_i \rangle} {\langle \Psi_T | \exp [ - H \tau ] | \Psi_i \rangle},
\label{eq:mixed}
\end{equation}
where $\Psi_i$ is an initial state used to start the simulation and $\Psi_T$ is a trial state incorporating as much knowledge as possible of the ground state.  It is often possible to take $\Psi_i = \Psi_T$, though this is not required. In the limit of large imaginary time $\tau$, the $O (\tau)$ can be used to
determine the ground-state properties of the system. The energy is the simplest: in that case the Hamiltonian  commutes
with the propagator $\exp [ -H \tau]$ and one can calculate the ground state expectation value. For other properties one must insert the propagation symmetrically
between the initial and final state or create a new Hamiltonian $H' = H + \epsilon O$, in either case  evaluating the expectation value of the operator:
\begin{equation}
\langle O (\tau,\tau') \rangle = \frac{\langle \Psi_T | \exp [ -H \tau ] O \exp [ - H \tau' ] | \Psi_i \rangle} {\langle \Psi_T | \exp [ - H (\tau + \tau ') ] | \Psi_i \rangle},
\end{equation}
which in the limit of large $\tau$ and $\tau'$ gives the true ground state expectation value.

\section{Equation of State}

   The equations of state for cold atoms and for neutron matter have been extensively studied
theoretically, and the cold atom system has been extensively studied experimentally as well.
The cold atom system is very simple: for a zero-range interaction the equation of state
is a function of only the product of the Fermi momentum and the scattering length ($k_F a$).
More specifically, the energy can be written as a function of $k_F a$ times the energy of
a non-interacting Fermi Gas ($E_{FG} = ( 3/5) (\hbar^2 / 2m) k_F^2$) at the
same density. Here the Fermi momentum 
is defined through the density of the corresponding non-interacting gas:
$k_F = (3 \pi^2 \rho)^{1/3}$.
At unitarity (infinite scattering length) the ratio
of the energies of interacting and noninteracting Fermi gases
 $E/E_{FG}$ is typically called the Bertsch 
parameter $\xi$.\cite{bertsch1998}

\subsection{Unitarity}

\begin{figure}[ht]
\centerline{\includegraphics[width=0.8\columnwidth]{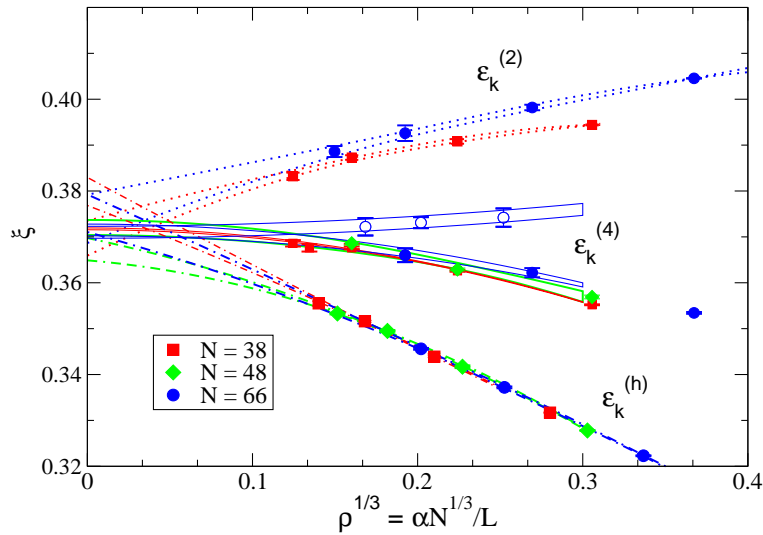}}
\caption{AFMC lattice calculations of the unitary Fermi Gas $\xi$ parameter, updated from Ref. \citenum{carlson2011auxiliary}.  Symbols are for different kinetic terms as a function of particle number and lattice size. The lattice spacing
is denoted as $\alpha$.
Simulations have been performed with $L^3$ lattices, for different values of 
lattice length $L$ in each direction; 
open symbols are for even L=16,20,24; closed are for odd L (see text).  All extrapolations are consistent with
$\xi = 0.372 (5)$.}
\label{fig:latticeunitarity}
\end{figure}

A history of results for the Bertsch parameter is given in Ref. \citenum{endres2012}.  The first
DMC calculation used up to 40 particles and a modified Poeschl-Teller potential with $k_F r_e \approx 0.3$,
where $r_e$ is the effective range of the interaction,
and yielded a fixed-node energy of $\xi = 0.44(1)$.\cite{carlson2003superfluid}
Subsequent DMC calculations used improved trial functions, larger particle numbers, and
better extrapolations to $k_F r_e \rightarrow 0$ to yield $\xi = 0.40(1)$.\cite{carlson2008superfluid}
The best present DMC result is from the calculations of Ref. \citenum{forbes2011resonantly},
while an updated extrapolation to $r_e \rightarrow 0$ gives $\xi = 0.390(1)$ \cite{gandolfi2011bec} for
an upper bound.  This calculation also carefully compared results at finite particle number to
a superfluid Local Density Approximation (LDA) to extrapolate to large N. It was found that calculations for $N=38$
or larger are very close to the thermodynamic limit.

There is also a substantial history of lattice simulations, both for the ground-state,\cite{lee2006,lee2006ground,lee2007superfluidity,lee2008ground,abe2009from}
and at finite temperature.\cite{bulgac2006,bulgac2008quantum} The earliest ground-state calculations
estimated $\xi = 0.25(3)$, for systems up to 22 particles on lattices up to $6^3$.
The recent calculations of Ref. \citenum{carlson2011auxiliary} use branching random walks and
a BCS trial function and importance sampling for systems of 66 particles
on lattices up to $27^3$ and obtain $\xi = 0.372(5)$ for several different actions.  Updated results for
these calculations are shown in Figure \ref{fig:latticeunitarity}.

In the figure, the upper curves use a $k^2$ dispersion relation tuned to unitarity. This
$k^2$ dispersion has a 
finite positive effective range of $0.337 \ \alpha$, where $\alpha$ is the lattice spacing. The middle set
of curves adopt a $k^2 + k^4$ dispersion that is tuned to zero effective range, and the lower curves
use a Hubbard dispersion relation, which has a negative effective range of $-0.306\  \alpha$.
The $k^2 + k^4$ results show a set of simulations with even L as open symbols, while
simulations at odd L are shown as filled symbols.  The two sets of results are slightly displaced;
similar displacements have been found with limited statistics for the other dispersions.
All extrapolate to the same value of $\xi$ within statistical errors; we return to the dependence on
effective range below. A new lattice calculation in Ref. \citenum{endres2012} reports a higher value
of $\xi$, above the upper bound found in DMC calculations.

There have also been a large number of experimental determinations of $\xi$:
the original measurements\cite{bartenstein2004collective,kinast2005heat,partridge2006pairing} have found
qualitative agreement with the DMC calculations listed above.
More precise recent experiments have found $\xi = 0.39(2) $\cite{luo2009thermodynamic} and $\xi = 0.41(1)$\cite{navon2010equation}
with a smaller value of $\xi = 0.375(5)$ found most recently.\cite{ku2012revealing}
This experimental value is quite precise and overlaps our lattice results.

\subsection{Equation of State: Cold Atoms and Neutron Matter}

\begin{figure}[h]
\centerline{\includegraphics[width=0.8\columnwidth]{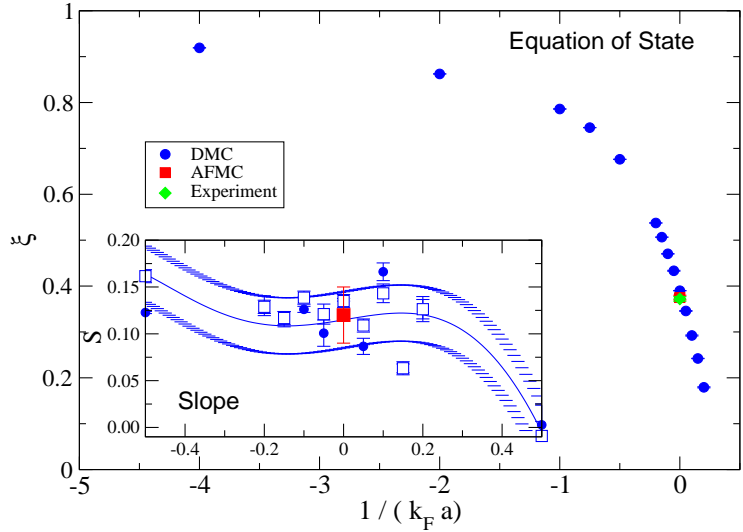}}
\caption{Equation of state of cold atoms versus $1/ ( k_F a)$. Blue circles are DMC calculations,
the red square and green diamond are lattice and experimental values at unitarity $1/ (k_F a)$ = 0.
The insert shows the corrections from finite effective range near unitarity (see text).}
\label{fig:coldatomeos}
\end{figure}

Of course the full equation of state ($E/ E_{FG}$) as a function of $k_F a$ is required 
to compare with neutron matter, which has a fixed, large effective range and must be studied
by varying the density.   The most recent DMC results for the full equation of state are presented
in Fig. \ref{fig:coldatomeos}, and compared to the lattice results and the most recent experimental
result.  These results are quite smooth as a function of $k_F a$ and extrapolate correctly in both
the BCS and BEC regimes.  

Because the cold-atom interaction is short-ranged, the derivative of the energy with respect
to $k_F a$ is given completely by short-range physics, as originally written down by Tan in a
series of papers.\cite{tan2008generalized, tan2008degenerate, tan2008energetics}
The derivative of the energy per particle with respect to $k_F a$ is given, using the Hellman-Feynman
theorem, by:
\begin{equation}
\frac{dE}{da^{-1}} \ = \ \frac{N}{2} \int d^3 \bfr  g_{\uparrow\downarrow} (r) \frac{d V(r)}{d a^{-1}}
\end{equation}
The pair distribution $g_{\uparrow\downarrow} (r) \rightarrow 0 $ goes like $A^2/r^2$ at unitarity for
small $r$, with $g_{\uparrow\downarrow} (r) \rightarrow 1/2 $ at large r. The change 
in energy with respect to $a^{-1}$ is
\begin{equation}
\frac{d E}{d a^{-1}} = \ - \ \frac{\hbar^2 2 \pi \rho A^2}{m} \rightarrow C = 8 \pi^2 \rho^2 A^2,
\end{equation}
where $C$ is Tan's contact parameter.  Near unitarity the EOS is conventionally parametrized as
\begin{equation}
\frac{E}{E_{FG}} \ = \ \xi - \frac{\zeta}{k_F a} + ...,
\end{equation}
with $ \zeta = (5 \pi / 2) C / k_F^4$. We return to the contact parameter in the discussion of
short-range physics below.

\begin{figure}[h]
\centerline{\includegraphics[width=0.8\columnwidth]{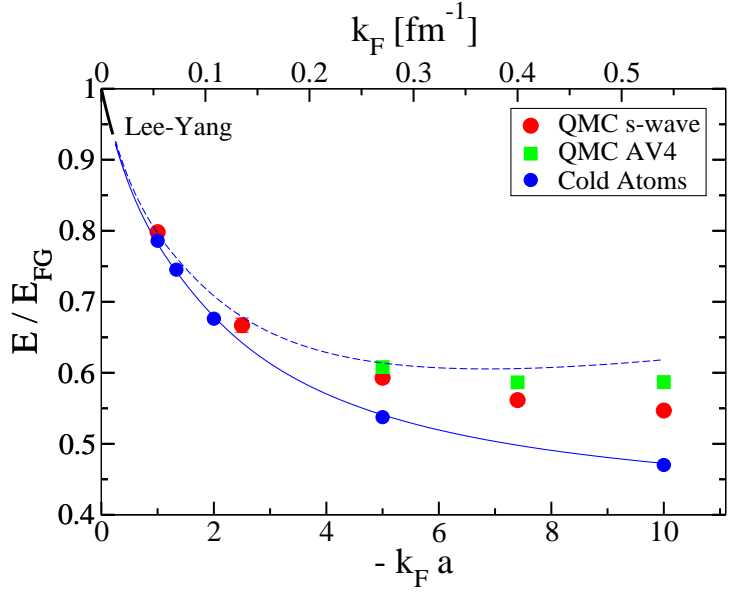}}
\caption{Comparison of the equation of state of cold atoms and neutron matter at low density.
Neutron matter calculations are from Ref. \citenum{gezerlis2010low}.  Differences at low density
are primarily due to the effective range of the neutron-neutron interaction. The solid line is a 
fit to the cold atom results, the dashed line includes an estimate of effective range effects (see text).}
\label{fig:nmatteratomeos}
\end{figure}

In Fig. \ref{fig:nmatteratomeos} these cold atom results are compared to the QMC for neutron matter,\cite{gezerlis2010low} and to the analytic expression available at small $k_F a$. At low densities, the
neutron matter and cold atom results agree, they also agree with a simple extrapolation 
of the analytic results near $k_F a = 0$.  At higher densities, the cold atom and neutron
matter equations of state start to diverge somewhat
as the effective range becomes important.  The dependence of the equation of state on effective
range can be made explicit, as we discuss below.  This dependence gives a quantitative picture
of the difference between neutron matter and cold atoms that could perhaps be tested in cold atom experiments with narrow
resonances. We will return to the finite-range corrections below.

\subsection{Equation of State: Unequal Masses}

Cold atom experiments can also be performed with species of different mass, providing
important information about the structure of the ground state of the unitary Fermi Gas.
For species of different mass $m_\uparrow$ and $m_\downarrow$,  if we normalize the $\xi$ parameter
by the reduced mass $E_{FG} = \frac{\hbar^2 k_F^2}{4 \mu} $, BCS theory would give a
value of $\xi$ independent of the mass ratio $r = m_\uparrow / m_\downarrow$.
The difference in Hamiltonians for equal $(r=1, m_\uparrow = m_\downarrow = m) $
masses and unequal masses is
\begin{eqnarray}
\Delta H & = & \
\sum_{i=1}^{N_\uparrow}  - \ \frac{\hbar^2 \nabla_i^2}{2 m_\uparrow} +
\sum_{i=1}^{N_\downarrow}  - \ \frac{\hbar^2 \nabla_i^2}{2 m_\downarrow} -
\sum_{i=1}^{N_\uparrow+N_\downarrow}  - \ \frac{\hbar^2 \nabla_i^2}{2 m} \nonumber \\
& = &   \sum_{i=1}^{\# pairs}  - \frac{ \nabla_i^2}{4 m} \frac{(r-1)^2}{(r+1)^2},
\label{eq:unequalmass}
\end{eqnarray}
where in the last line the particles have arbitrarily been divided into $N/2$ spin up - spin down
pairs.

\begin{figure}[h]
\centerline{\includegraphics[width=0.8\columnwidth]{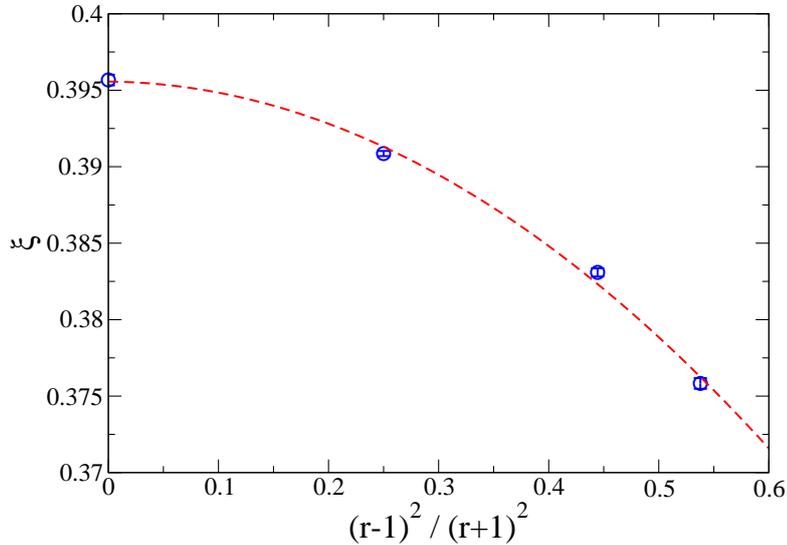}}
\caption{Dependence of the unitary Fermi Gas equation of state on mass ratio for 
fixed reduced mass $\mu = m_\uparrow m_\downarrow / (m_\uparrow + m_\downarrow)$,
plotted versus $(r-1)^2 / (r+1)^2$, where $r$ is the mass ratio.}
\label{fig:gandolfimassratio}
\end{figure}

Figure \ref{fig:gandolfimassratio} shows the DMC calculations of $\xi$ for different mass
ratios.  Initial calculations for different mass ratios were reported in Ref. \citenum{gezerlis2009heavy}.
From Eq. \ref{eq:unequalmass} we can see that the energy change can be evaluated
in perturbation theory near r=1.
\begin{equation}
\Delta (E/N) \ =  \ \langle 0 | \Delta H | \ 0 \rangle = (1/2) \langle P_{ij}^2/(4m) \rangle |_{r=1}
\frac{(r-1)^2}{(r+1)^2},
\end{equation}
where the 1/2 comes from the number of pairs (N/2), $P_{ij}$ is the total momentum of a pair,
and the expectation value is to be taken in the ground state of the equal mass unitary gas.
Note that these calculations were performed for small but finite value of the effective range, yielding
a slightly larger value of $\xi$ than at zero effective range.
This rather asymmetric way of writing the energy difference is valuable because it tells us
something about the character of the state.  For a BCS-like state with all pairs at $P=0$
the energy difference is zero in first-order perturbation theory.  Of course the free Fermi Gas
can also be written in this manner.  The difference is finite for the case when the ground
state wave function does not have a spin down particle at $-{\bf p}$ for every spin up particle
at momentum ${\bf p}$.

Fig. \ref{fig:gandolfimassratio} shows the DMC calculations as points with error bars,
and a quadratic fit to the data.  The linear coefficient in this fit is very small, consistent
with zero within statistical errors.  Thus to a very good approximation the ground state
of the unitary gas can be written as a state of pairs with zero momentum.  To
confirm this result it would be
important to have experimental measurements of the energy for several different
mass ratios.

\subsection{Equation of State: effective range}

\begin{figure}[h]
\centerline{\includegraphics[width=0.8\columnwidth]{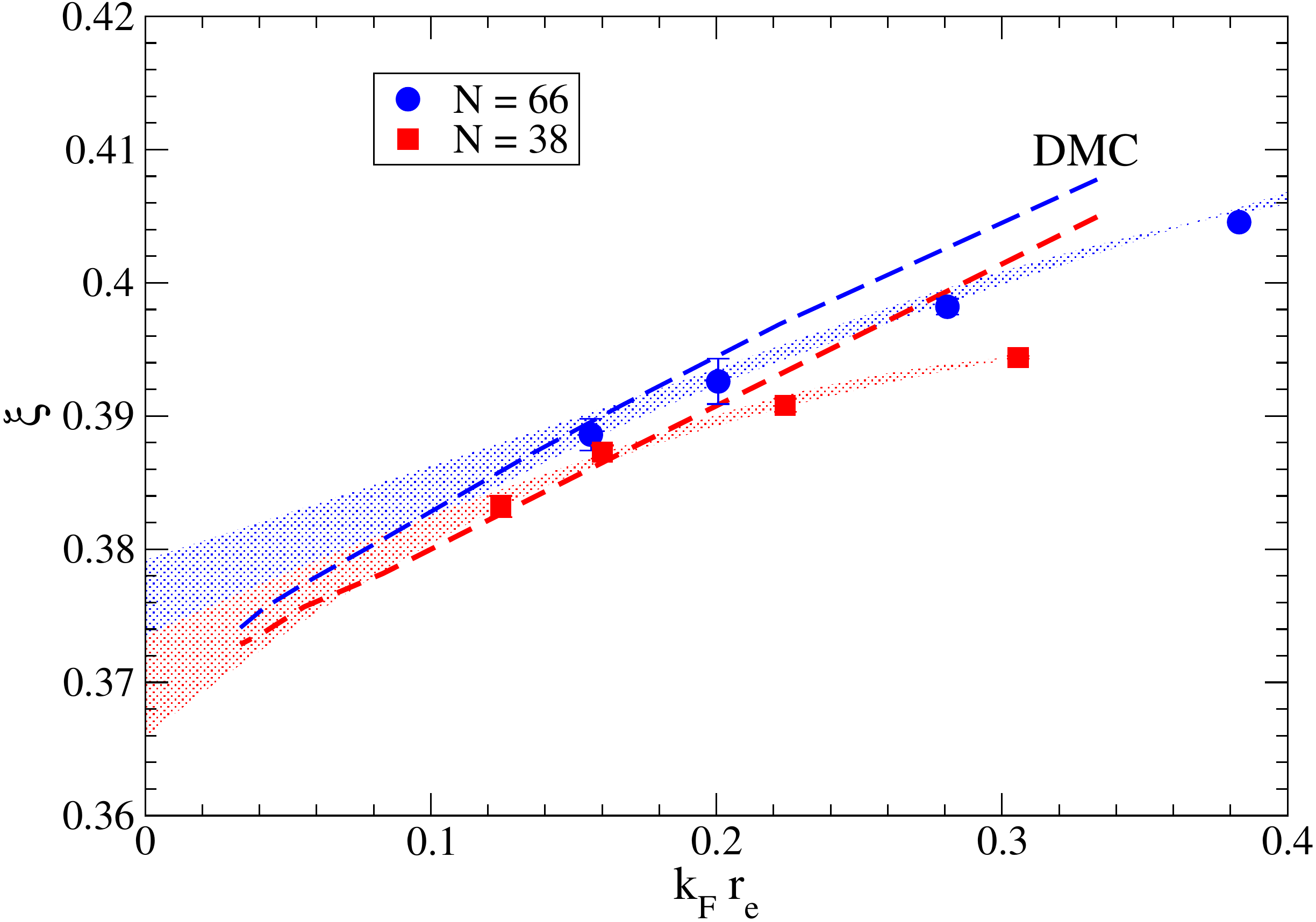}}
\caption{Dependence of the unitary Fermi Gas equation of state on Fermi momentum
vs. effective range $(k_F r_e)$.  Shaded bands  are fits to the lattice results, and
dashed lines give DMC results.}
\label{fig:eosvsre}
\end{figure}

As is apparent from Fig. \ref{fig:latticeunitarity}, the equation of state for
finite effective range $r_e$ varies  linearly with $k_F r_e$ at small effective range.
In Ref. \citenum{carlson2011auxiliary}, the equation of state at unitarity for different
effective ranges was found to be:
\begin{equation}
\xi ( k_F r_e ) = \xi (0) + S k_F r_e,
\label{eq:reff}
\end{equation}
where $\xi(0)$ characterizes the ground-state energy at zero effective range,
and $S$ is the slope parameter giving the linear dependence on $k_F r_e$.
The slope parameter was extracted from both DMC and AFMC calculations,\cite{carlson2011auxiliary}
and found to be $S = 0.12(3)$.  The results for different effective ranges are
shown in Figure \ref{fig:eosvsre}.  More recent DMC results for a variety of
interactions have recently appeared,\cite{forbes2012effective} they find $S = 0.127(4)$
using a variety of interactions. These calculations further demonstrate that $S$ is
a universal parameter, as originally conjectured in the original version of Ref. \citenum{werner2012}.
In this manuscript, the authors also claim that lattice results will in general have a dependence
on the total momentum $P$ of a pair.  For the unitary gas, however, the expectation value
of $\langle P^2 \rangle$ is approximately zero, as shown above.  Therefore the lattice 
and continuum results are both in agreement with Eq. \ref{eq:reff}.

In DMC calculations the slope parameter $S$ is not too sensitive to $k_F a$ near unitarity.
Fig. \ref{fig:coldatomeos} shows, in the inset, the slope parameter $S$ evaluated 
from DMC calculations near unitarity.  It is positive and approximately $0.1$ near unitarity,
but changes significantly in the BCS and BEC regimes.
The difference between the cold atom EOS and neutron matter at sufficiently small
densities should be approximately $\xi_{neutrons} - \xi_{atoms} \approx S k_F r_e$,
or approximately 0.05 at $- k_F a = 5$ since the neutron-neutron
effective range is expected to be approximately 2.7 fm. Fig. \ref{fig:nmatteratomeos}
shows a fit to the cold atom results at zero effective range as a solid line.  The dashed
line adds an effective range correction with $S\ =\ 0.1$.  This should be the dominant
correction at $ \ k_F   \leq 0.25$ fm$^{-1}$ , near $  k_F  \approx 0.5$ fm$^{-1}$ one would
have $  k_F r_e \approx 1 $ and higher order corrections in $s-$ and $p-$wave interactions
could be important. 

\section{Pairing Gap}

Both low-density neutron matter and cold atoms are strongly paired Fermi systems,
they exhibit some of the largest pairing gaps of any systems known when measured
in terms of the Fermi energy.  We define the pairing gap at T=0 as the difference
between the energy of an odd particle system and the average of the two nearby even particle
systems in periodic boundary conditions:
\begin{equation}
\Delta \ = \ E(N+1) - ( E(N) + E(N+2))/2,
\end{equation}
with the universal parameter $\delta$ defined as the pairing gap divided by the Fermi energy
$E_F = \hbar^2 k_F^2 / 2m$.
For simulations of a large enough number
of particles this should correspond to the traditional definition of the pairing gap.

The fact that the pairing gap is so large, a sizable fraction of the Fermi energy,
makes it possible to use QMC methods to accurately calculate the gap
by separately calculating the energies of the even and odd particle systems. \ In addition,
the fact that the energy per particle shows no significant shell effects for reasonably
small systems $( N > 30)$ makes it much easier to approach the continuum limit.
Though there is an upper bound principle for the even and odd systems, there
is no specific bound on the pairing gap. 

The original calculations of the pairing gap in cold atoms at unitarity found $\Delta  / E_{FG}
\approx 0.9$ or $\delta = 0.55 (5)$.\cite{carlson2003superfluid}  Subsequent improvements
to the wave function\cite{carlson2005asymmetric} found a slightly reduced value for
the gap, $\delta = 0.50(5)$. These results can be compared to an extraction of the
pairing gap from the measured density distributions in partially spin-polarized 
trapped cold atoms\cite{carlson2008superfluid} and measurements of the RF response 
in such systems\cite{schirotzek2008determination}, who find $\delta = 0.45(5)$ and $\delta=0.44(3)$,
respectively.

\begin{figure}[h]
\centerline{\includegraphics[width=0.8\columnwidth]{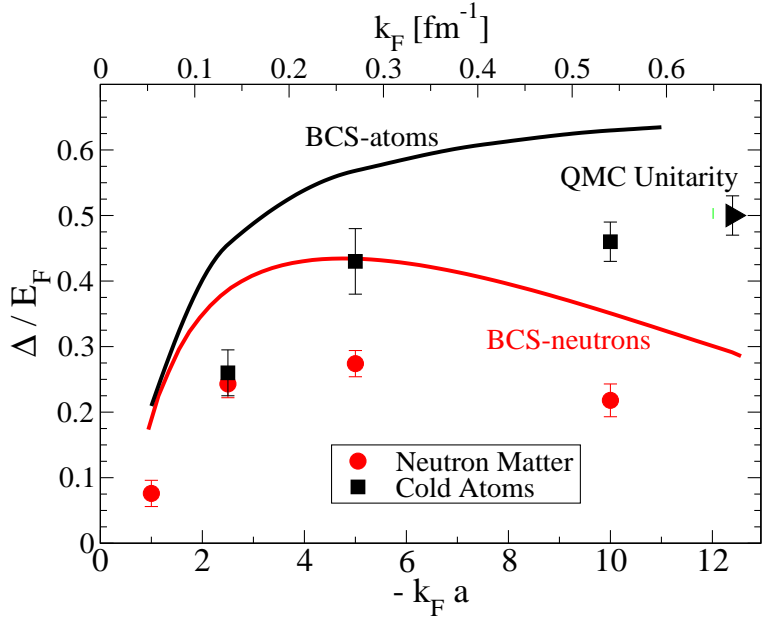}}
\caption{Pairing gap in cold atoms and neutron matter. BCS mean-field results
are shown as solid lines, DMC results are shown as symbols.
The pairing gaps are divided with the relevant one-body quantity, namely the
Fermi energy $E_F$ (analogously to the ground-state energy in Fig.~\ref{fig:neutronatomdrops}
being divided with $E_{FG} = 3E_F/5$.}
\label{fig:gapoveref}
\end{figure}

The pairing gap in neutron matter has historically been the subject of a great deal
of interest and theoretical activity.\cite{Gandolfi:2008,Gandolfi:2009}
QMC calculations of the pairing gap were
performed in \citenum{gezerlis2008strongly} and \citenum{gezerlis2010low}.  These calculations
used the $s$-wave and $s$- \& $p$-wave components of the AV18 interaction, respectively.
A summary of the results are shown in Fig. \ref{fig:gapoveref}.\cite{gezerlis2008strongly}

In the figure, BCS results are given by solid lines. In the weak-coupling limit,
the pairing gap is expected to be reduced from the BCS value by $(1/4e)^{1/3} \approx 0.45$
from the Gorkov polarization correction.\cite{gorkov1961}  It is difficult for QMC calculations to
calculate the pairing gap at coupling weaker than  $k_F a = -1$ because of the delicate cancellations.
At this coupling, though, we find a suppression in the gap roughly compatible with the Gorkov
suppression.  At stronger coupling the suppression diminishes smoothly, and the gap reaches
a value of $0.50(05)$ at unitarity.  In the BEC regime the pairing gap approaches half the binding
energy of the pair as reproduced by the BCS equation.

The calculated gaps in neutron matter are considerably smaller than in cold atoms, but still
reach a maximum of nearly $0.3 E_F$ at $k_F a = -5$.  The effective range in the neutron-neutron
interaction reduces the gap significantly, as shown in the comparison of BCS results and in the
Monte Carlo calculations. These pairing gaps are considerably larger than found in many
diagrammatic approaches,\cite{gezerlis2010low} but in agreement with the lattice results
of Ref. \citenum{abe2009from}.

\section{Short-range physics}

One can also investigate the short-range (high-momentum)  physics in cold atom experiments.
Because of the simple short-range interaction, this short range physics is directly related to the
equation of state discussed above.  The pair distribution function and off-diagonal density matrix
at short distances and the  momentum distribution at high momenta are all governed by the contact parameter.

\begin{figure}[h]
\centerline{\includegraphics[width=0.8\columnwidth]{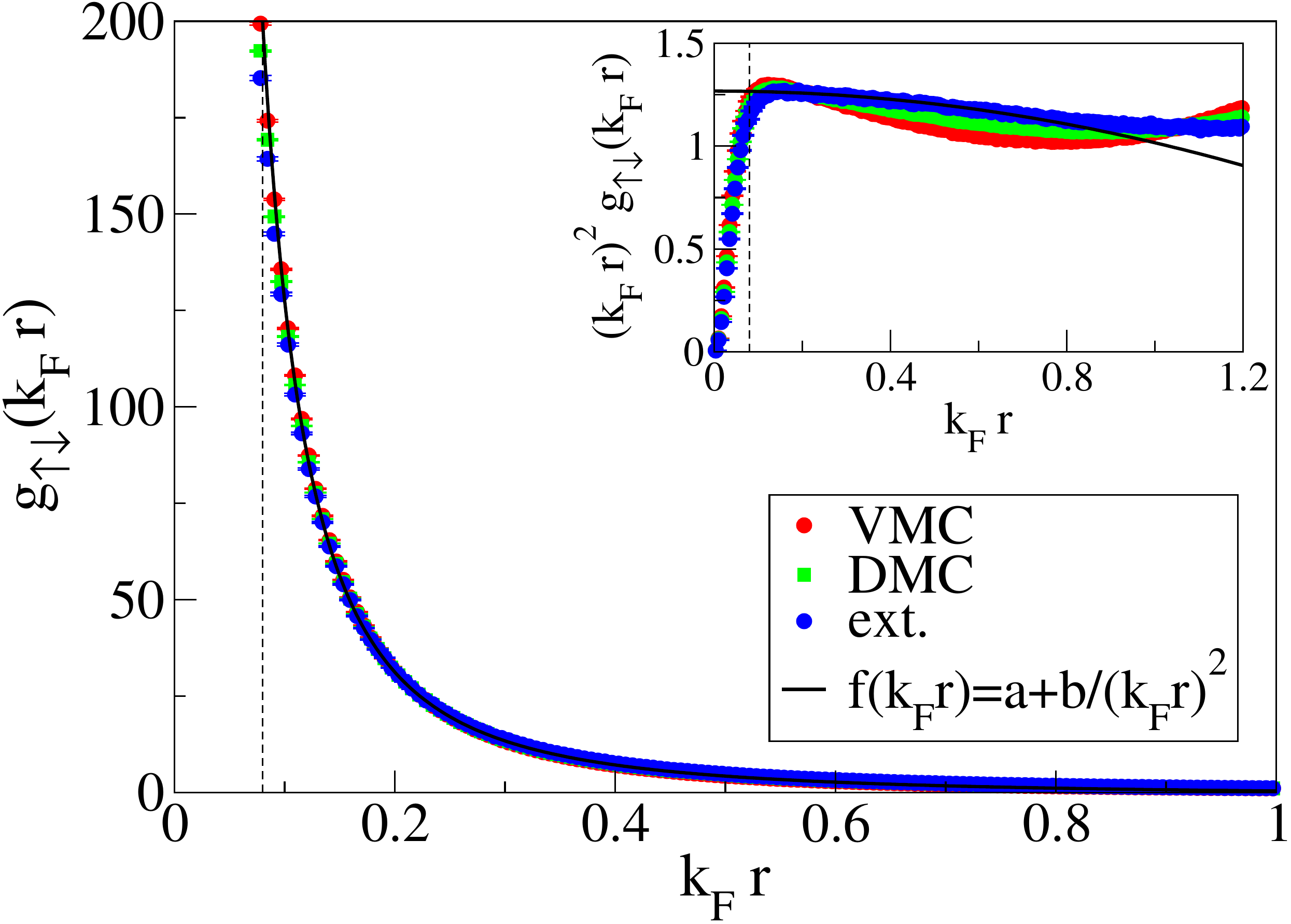}}
\caption{Pair distribution function $g_{\uparrow\downarrow} (r)$ for cold atoms at unitarity.
Inset shows the behavior at short distances scaled by $r^2$, the magnitude of the contact determines
the value of this quantity (see text). }
\label{fig:gofr}
\end{figure}

The pair distribution function is shown in Fig. \ref{fig:gofr}; the contact governs the huge spike
near $r=0$. In the inset the pair distribution function is multiplied by $(k_F r)^2$ to show the
behavior near $r=0$, the dip at very short distances is due to the finite range interaction used in
the simulations. The behavior in the region of the vertical dashed line and slightly beyond is
governed by universal physics. In the figure the red points are variational Monte Carlo (VMC) 
results obtained from the trial wave function $| \Psi_T \rangle$, the green are the mixed estimates
of the form obtained from Eq. \ref{eq:mixed} and blue are the
full extrapolated DMC results, as is apparent in the figure the VMC calculation is correctly capturing the basic physics.

\begin{figure}[h]
\centerline{\includegraphics[width=0.8\columnwidth]{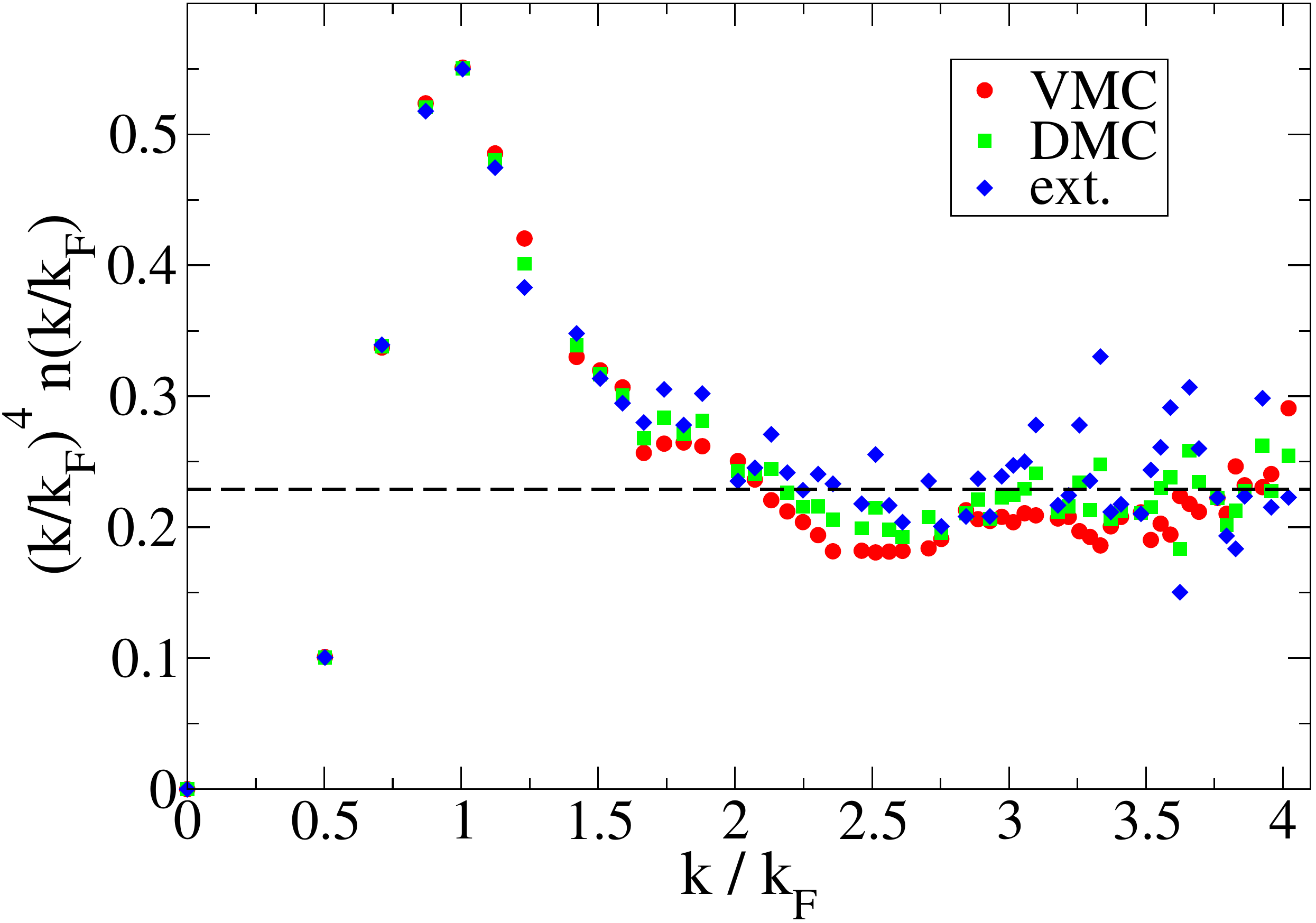}}
\caption{Momentum distribution scaled by $k^4$ for cold atoms at unitarity.}
\label{fig:nofk}
\end{figure}

The momentum distribution scaled by $k^4$ is plotted in Fig. \ref{fig:nofk}. The momentum distribution
at large $k$ is proportional to the contact.  The horizontal line in the figure is the value that would be
expected from calculations of the equation of state: $\zeta = 0.901(2)$.  Extractions of the contact from
all these observables are consistent with this value, though of course some are noisier than others.
It will be interesting to see what information on the short-range physics of neutron matter can be
obtained from theory and experiments with narrow resonances with a significant effective range.

Initial experiments on the spin and density response of cold atoms have also been performed.\cite{hoinka2012dynamic}
These and future results will be very interesting as they can tell us about the propagation of
particles and spin in the unitary gas. The response functions can be written as:
\begin{eqnarray}
S_{\rho} (q, \omega) \ & = &  \sum_f \ \langle 0 | \sum_i \exp [- i {\bf q} \cdot \bfr_i ] | f \ \rangle 
\langle f | \sum_j \exp [ i {\bf q} \cdot {\bfr_j}] | 0 \rangle \ \delta ( \omega - (E_f - E_0)) \nonumber \\
S_{\sigma} (q, \omega) \ & = &  \sum_f \ \langle 0 | \sum_i \exp [- i {\bf q} \cdot \bfr_i ] \ {\mbox{\boldmath$\sigma$}}_i\  | f \ \rangle \cdot
\langle f | \sum_j \exp [ i {\bf q} \cdot {\bfr_j} ] \ {\mbox{\boldmath$\sigma$}}_j \  | 0 \rangle \ \delta ( \omega - (E_f - E_0)) \nonumber \\
\end{eqnarray}

These response functions have been calculated at high momenta in terms of the operator production 
and related high-momentum expansions.\cite{son2010short,goldberger2012structure,hofmann2011current,hu2012universal,nishida2012probing}  The experiments show a two-peak structure in the density response, one
at $\omega = q^2 / (2m)$ associated with the breaking of a pair and one at $\omega = q^2/(4 m)$ associated
with the propagation of a pair.  The spin response requires breaking of a pair.  The initial experiments are
at rather high momentum transfer, many times the Fermi momentum, and hence probe the short-range physics.
It will be interesting to see how these response functions evolve at lower momenta.
One can also compare calculations and experiments on the sum rules associated with the contact
parameter by integrating the response over $\omega$.

\section{Inhomogeneous Matter}

Finally, we turn to the properties of inhomogeneous matter. This is the regime with perhaps the
closest connection between nuclear physics and cold atom physics.  In the inner crust of a neutron star
the neutrons form a gas between the neutron-rich nuclei that exist in a lattice structure. This neutron
matter is very low density and is inhomogeneous, and for many of the transport properties we would
like to understand the behavior of this gas.  The properties of inhomogeneous neutron matter, particularly
the gradient terms, are very difficult to determine from the binding energies of atomic nuclei.
The isovector gradient term is one of the least constrained parameters in nuclear density functionals,
and ab-initio calculations can provide valuable guidance.

\subsection{Inhomogeneous Matter: Bulk Properties}

We have calculated the properties of finite systems of neutrons bound in harmonic and Woods-Saxon wells.\cite{gandolfi2010cold}
Original calculations of these drops\cite{pudliner1996neutron} were limited to N=8 neutrons because of 
the spin-dependence of the nuclear interaction.  These more recent calculations use Auxiliary Field Diffusion Monte Carlo\cite{sarsa2003neutron,gandolfi2007quantum,gandolfi2009quantum} methods and Green's function Monte Carlo methods; the former have been used to treat
up to N=50 neutrons.  The AV18 NN interaction plus the UIX three-nucleon interaction have been used for
these calculations. As these wells produce fairly modest densities for 50 particles or less, the three-neutron
interaction is not very important.

\begin{figure}[h]
\centerline{\includegraphics[width=0.8\columnwidth]{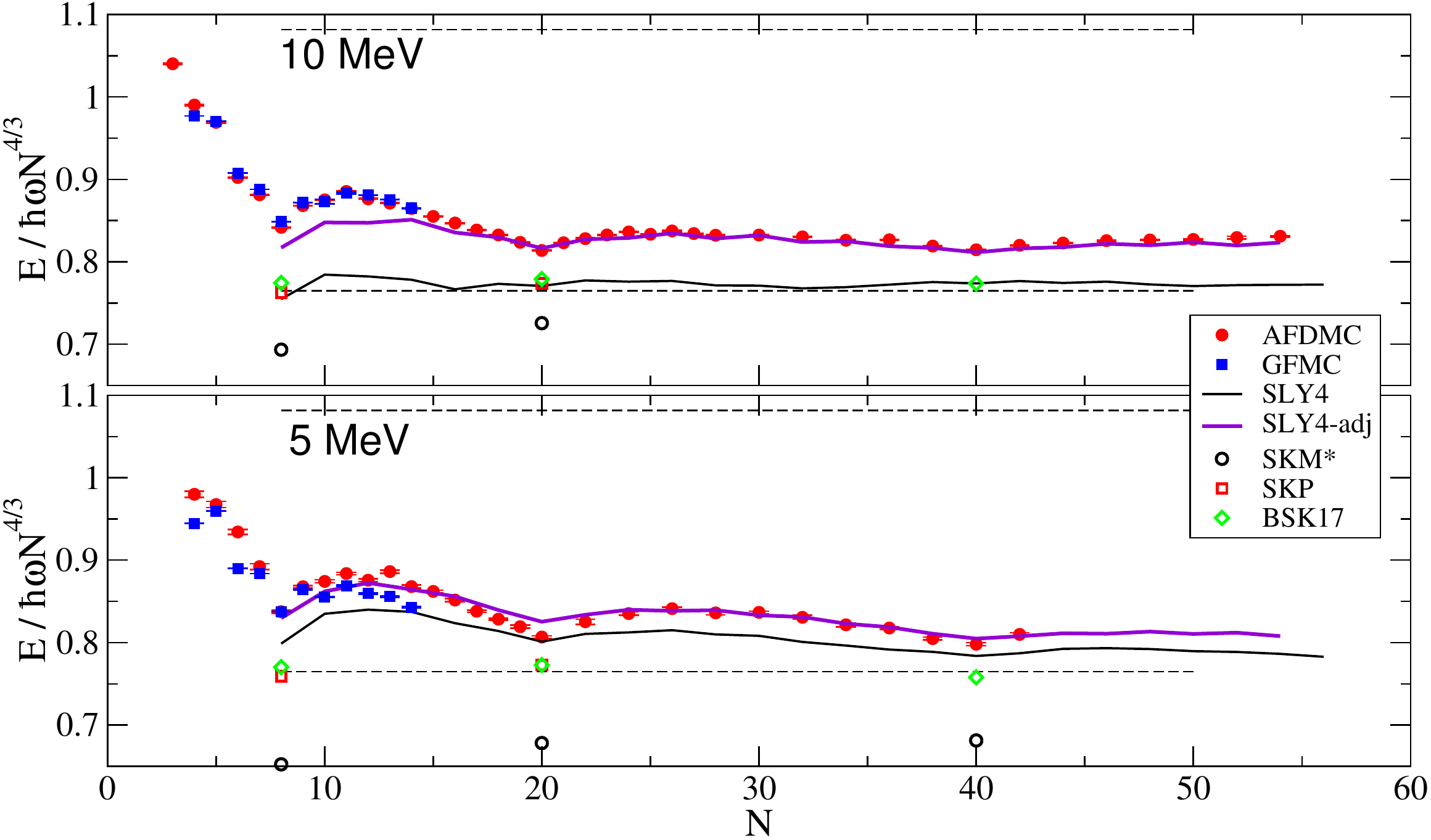}}
\caption{Scaled energies for neutrons bound in a harmonic well.\cite{gandolfi2010cold} 
The upper and lower straight dashed lines
are the Thomas-Fermi mean-field results for free fermions and for a scaled EOS $E/N = \xi E_{FG}$
with $\xi = 0.5$.  The open symbols are calculations with previous generation density functionals, and the
filled symbols are GFMC and AFDMC calculations.  The jagged line shows the results for the $SLy4$
density functional. The upper panel is for a harmonic trap frequency of 10 MeV and the lower for 5 MeV.}
\label{fig:ndrop}
\end{figure}

The energies in Fig.~\ref{fig:ndrop} have been scaled by $1 / (\hbar \omega N^{4/3})$, the expected behavior
in the Thomas-Fermi approximation for an EOS of the form $E/E_{FG} = \xi$.  The upper horizontal dashed lines
are for $\xi = 1$ (free fermions), and the lower for $\xi = 0.5$.  Neutron matter over a considerable
range of densities is roughly consistent with $\xi = 0.5$, though it is less attractive at low and high densities.
The traps have harmonic frequencies of $\omega = 10$ MeV (upper panel) and $5$ MeV ( lower panel).
Results for several typical older-generation density functionals are shown as open circles, and calculations
using SLY4 as solid lines.

The Skyrme interactions typically give significantly lower energies than the microscopic calculations,
particularly for the 10 MeV well. The greater difference for the 10 MeV well suggests that isovector gradient
terms in the density functional should be more repulsive. The curves marked ``SLY4-adj'' in the figure are
obtained by adjusting the isovector gradient, the isovector pairing, and the isovector spin-orbit terms in the
interaction.  At the closed shells ($N=8,20,40$) only the change in the isovector gradient term is important.
A reasonable fit can be obtained to the closed-shell energies in the 5 and 10 MeV harmonic wells and the Woods-Saxon wells with a single adjustment to the isovector gradient term. This adjustment also better reproduces
the rms radii and mass distributions of the ab-initio calculations.\cite{gandolfi2010cold}

\begin{figure}[h]
\centerline{\includegraphics[width=0.8\columnwidth]{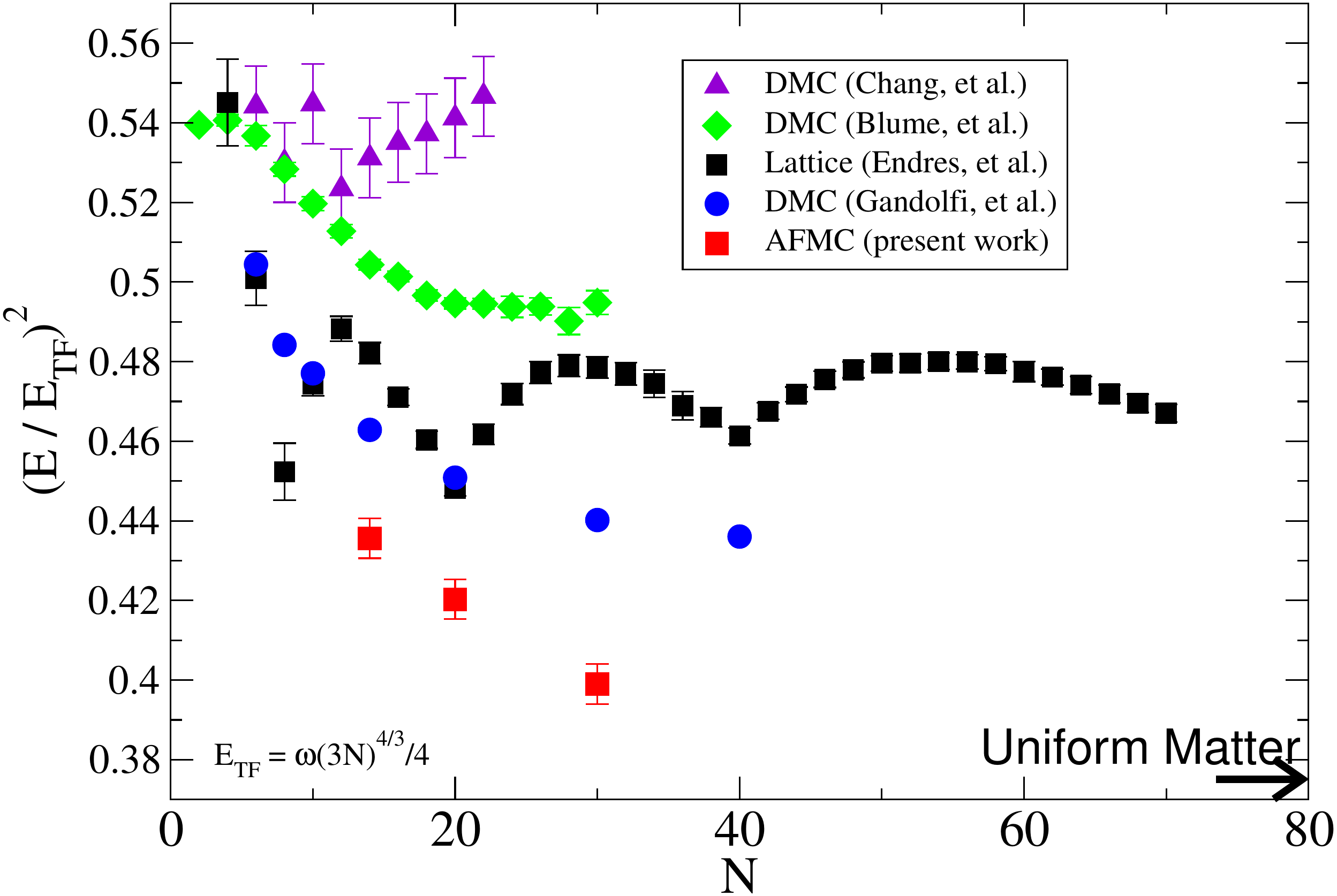}}
\caption{Comparison of different calculations of the harmonically confined unitary Fermi Gas.\cite{chang2007unitary,blume2007universal,endres2011,gandolfi2012prep,forbes2012effective}
The vertical axis is the square of the energy divided by Thomas-Fermi energy $E_{TF} = \omega(3 N)^{4/3} /4$,
the horizontal axis is the number of particles.  For a very large system $(E/ E_{TF})^2$ should
approach $\xi$.}
\label{fig:atomdrops}
\end{figure}

It is interesting to compare the neutron drop results to those obtained for cold atoms.
In Fig. \ref{fig:atomdrops} we plot the scaled energies of cold atom systems obtained in DMC
and AFMC calculations as well as the results of previous calculations.\cite{chang2007unitary,blume2007universal,endres2011}  
Two previous DMC calculations used fairly simple trial wave functions,\cite{chang2007unitary,blume2007universal}
the first using an orbital basis for the trial function and the second using a simple $1/r$
BCS pairing function as the trial wave function.  Both resulted in energies far above
what would be expected in local-density (Thomas Fermi) theory with $\xi \approx 0.4$.
New lattice results\cite{endres2011} yield somewhat lower energies, but significant shell
structure.  They are also above the energies expected from the measured and calculated values
of $\xi$ in the continuum, indicating either unnaturally large gradient corrections or other effects.

Our new DMC and AFMC calculations produce energies considerably lower than previous results.
The DMC energies are very smooth as a function of N as compared to neutron drop results, indicating
a lack of shell closures for the unitary Fermi gas with zero effective range.  These DMC calculations
use a more sophisticated trial function incorporating both a single-particle shell model basis and
substantial short-range pairing into the trial wave function. The resulting energy is somewhat
higher than the expectations from the bulk for N up to 50 particles.

It has been shown\cite{werner2006unitary} that the cold atoms trapped in harmonic wells have a
breathing mode associated with the scale invariance of the Hamiltonian of exactly $2 \ \hbar \omega$,
independent of particle number N, giving further evidence that cold atoms at unitarity
have a very weak shell structure, if any. Initial AFMC lattice calculations give energies
for 30 fermions much closer to the expected bulk limit, indicating a smooth and rapid transition
from few-particle systems to the bulk.  Further DMC and AFMC calculations are being pursued.

\begin{figure}[h]
\centerline{\includegraphics[width=0.8\columnwidth]{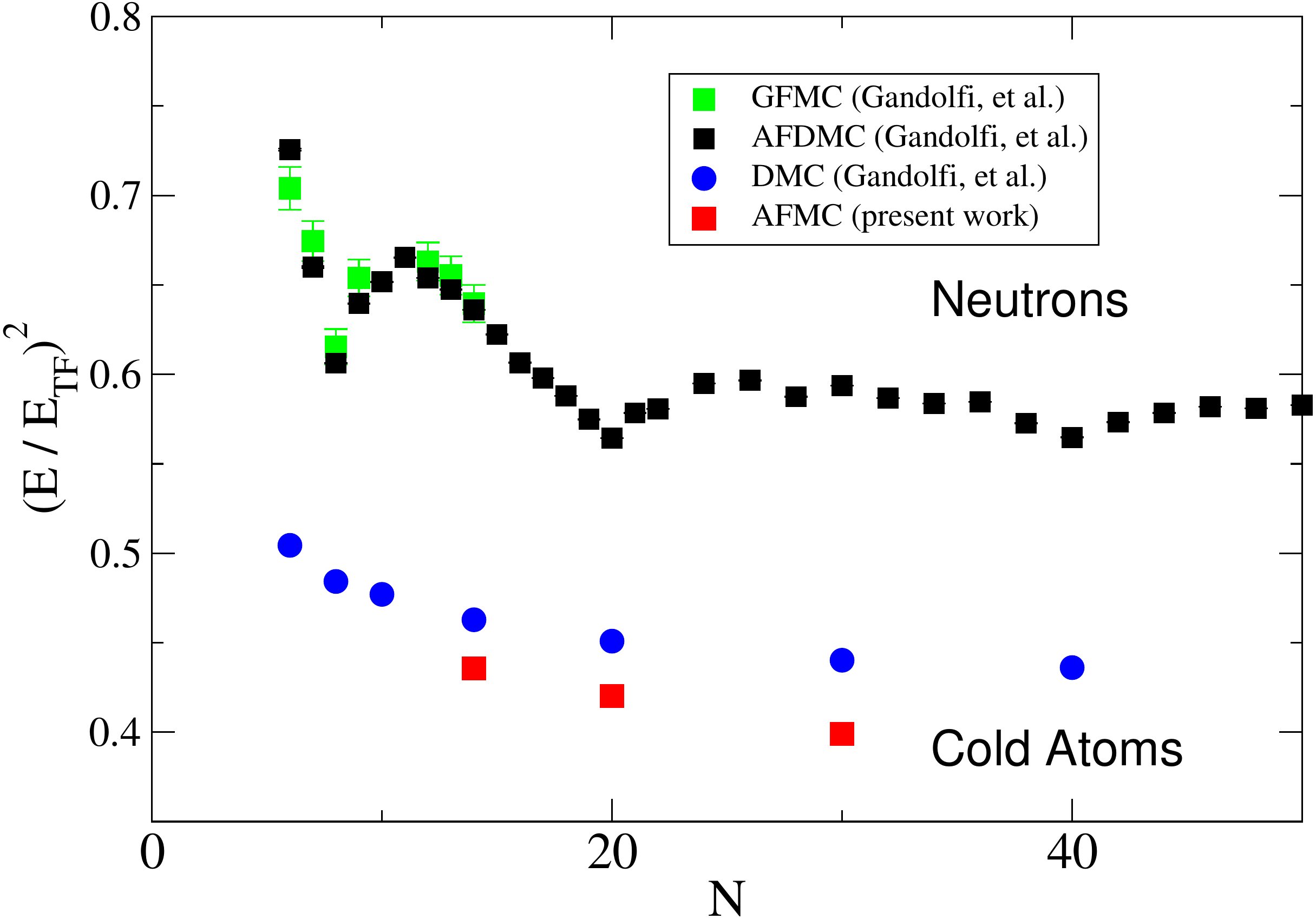}}
\caption{Comparison of harmonically confined neutron drops and cold atoms.  The
neutron drops have significant shell closures at N= 8, 20, 40, etc. because of the
finite effective range and further corrections.}
\label{fig:neutronatomdrops}
\end{figure}

In Fig. \ref{fig:neutronatomdrops} we compare cold atom results to neutron drops in the same
traps as a function of particle number  $N$.   
We plot the square of the energies because, in the local density approximation,
the square of the energy of the confined system is proportional to the energy
of uniform matter determined by $\xi$.  The bulk limit as obtained from the
lattice calculations shown in Fig. \ref{fig:latticeunitarity} is shown
as an arrow at the lower right of the figure.
The energies of neutron drops are considerably higher than those of cold atoms, this is at
least partially the result of the effective range in neutron matter. The gradient terms are likely
also important, however. These can be more precisely constrained by performing a local density
calculation using a realistic equation of state for neutron matter.

\subsection{Inhomogeneous Matter: pairing}

\begin{figure}[h]
\centerline{\includegraphics[width=0.8\columnwidth]{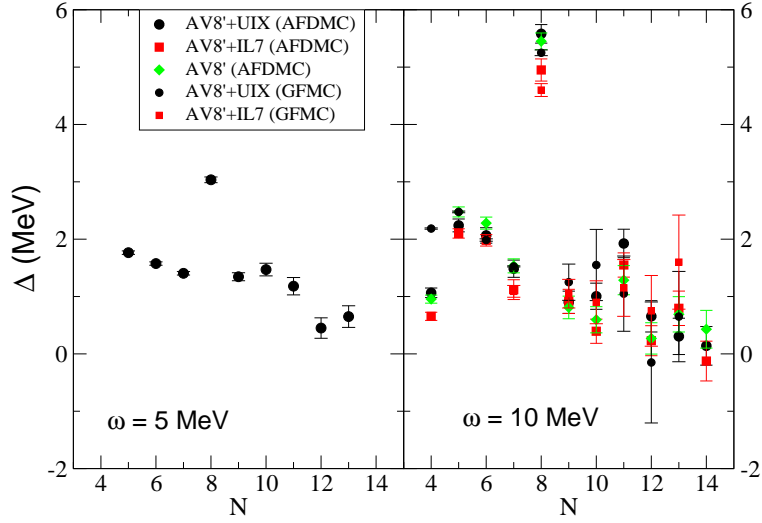}}
\caption{Calculated pairing gaps in harmonically confined neutron drops, at harmonic frequencies
$\omega = 5 $ MeV (left) and $10 $ MeV (right).  The large values at N $= 8$ are associated with the
shell closure. Gaps are calculated from the mass difference formula with a phase factor of $-1^{N+1}$
to take into the account the fact that the unpaired particles have a higher
energy associated with the pairing gap.}
\label{fig:ndroppairing}
\end{figure}

We have also performed calculations of the pairing gap in neutron drops, shown in 
Fig.~\ref{fig:ndroppairing}. In atomic nuclei
pairing is, at least predominantly, a bulk effect, in that the coherence length appears to be
comparable to the size of the nucleus.  This is to be expected in a regime where the effective
range is comparable to the interparticle spacing, in such cases it should be possible to
construct a mean-field theory that produces a qualitatively correct picture.

For cold atoms the conclusions will be quite different; for large enough systems the
unpaired atom will necessarily sit outside the center of the drop.  The gap is simply too high
in the high-density central region for the unpaired particle to penetrate. It will be very instructive
to compare theories and experiments as a function of scattering length, effective range,
and particle number.

\section{Conclusions}

Cold atom experiments and theory provide many valuable insights into our understanding
of strongly correlated fermions, and in particular have a close relationship with low-density
neutron matter. The equations of state for low-density neutron matter and cold atoms are
by now well understood and very similar as a function of $k_F a$, 
and the difference is understood in terms of a correction 
proportional to the fermi momentum times the effective range. 

Pairing gaps in cold atoms demonstrate a smooth transition from the BEC to BCS regime, and 
indicate that a sizable pairing gap is to be expected. Further experimental and theoretical studies
as a function of $k_F r_e$ at and near unitarity would be very valuable in providing explicit
confirmation and a more precise understanding.

A better understanding of the linear response of cold Fermi atoms could lead to new insights
into the dynamic response of neutron matter. It would be particularly valuable to map out
both the density and spin response of cold atoms as a function of momentum transfer.
Analogies to topics such as neutrino propagation in dense matter are clear, though there is
not a direct correspondence as there is for the equation of state and pairing gap.

Inhomogeneous matter is also quite intriguing, including small systems of trapped fermions,
fermions in optical lattices, and the transition from three to two-dimensions. Inhomogeneous
cold atom systems also have close analogies in nuclear physics, including the physics of nuclei
and the neutron star crust. Studies of narrow resonances with finite effective range could
help us understand the evolution of pairing from a local to a bulk phenomenon.
The rapidly expanding scope of cold atom experiments and Quantum Monte Carlo
will undoubtedly reveal intriguing new physics and close correlations with nuclear physics.

\section*{Acknowledgements}

The authors would like to thank Yusuke Nishida and Steven Pieper for valuable conversations and insights.
The authors are grateful to the LANL Institutional computing program, the NERSC computing facility
at Berkeley National Laboratory, and the Oak Ridge Leadership Computing Facility located in the Oak Ridge National Laboratory, which is supported by the Office of Science of the Department of Energy under Contract DE-AC05-00OR22725. The work of J. Carlson and Stefano Gandolfi is supported by the Nuclear Physics program
at the DOE Office of Science, and Stefano Gandolfi is also supported by the LANL LDRD program.
The work of Alexandros Gezerlis is supported by the Helmholtz Alliance Program
of the Helmholtz Association, contract HA216/EMMI ``Extremes of
Density and Temperature: Cosmic Matter in the Laboratory''.
  

\section*{Note}

This is a pre-copy-editing, author-produced PDF of an article accepted for publication 
in Progress of Theoretical and Experimental Physics following peer review. 
The definitive publisher-authenticated 
version Prog. Theor. Exp. Phys. 2012, 01A209 is available online 
\href{http://ptep.oxfordjournals.org/cgi/content/abstract/pts031?ijkey=4b4SmJxqyTqdYcH&keytype=ref}{here}.

\end{document}